# Supporting GNSS Baseband Using Smartphone IMU and Ultra-Tight Integration

Yiran Luo, You Li, Jin Wang, and Naser El-Sheimy

*Abstract*—A great surge in the development of global navigation satellite systems (GNSS) excavates the potential for prosperity in many state-of-the-art technologies, e.g., autonomous ground vehicle navigation. Nevertheless, the GNSS is vulnerable to various ground interferences, which significantly break down the continuity of the navigation system. Meanwhile, the GNSS-based next-generation navigation devices are being developed to be smaller, more low-cost, and lightweight, as the commercial market forecasts. This work aims to answer whether the smartphone inertial measurement unit (IMU) is sufficient to support the GNSS baseband. Thus, a cascaded ultra-tightly coupled GNSS/inertial navigation system (INS) technique, where consumer-level smartphone sensors are used, is applied to improve the baseband of GNSS software-defined radios (SDRs). A Doppler value is predicted based on an integrated extended Kalman filter (EKF) navigator where the pseudorange-state-based measurements of GNSS and INS are fused. It is used to assist numerically controlled oscillators (NCOs) in the GNSS baseband. Then, an ultra-tight integration platform is built with the upgraded GNSS SDR, of which baseband processing is integrated with INS mechanization. Finally, tracking and carrier-based positioning performances are assessed in the proposed platform for the smartphone-IMU-aided GNSS baseband via kinematic field tests. The experimental results prove that extra hardware with only a few dollars instead of more expensive ones can improve the GNSS baseband efficiently.

## I. INTRODUCTION

THE devices with the sensor receiving the global navigation satellite system (GNSS) signal are currently omnipresent all over the world. The GNSS offers all-weather, all-time, and high-precise absolute positioning results in civil, military, and industrial applications, such as autonomous ground vehicles [1], unmanned aerial vehicles (UAVs) [2], smartphones, and wearable sensors [3], [4]. Undoubtedly, the GNSS device will constantly exert its superiority and expand its influence in more fields in the future years. Nevertheless, the orbit of the GNSS satellite is very far from the surface of the Earth, and this fact makes received signals on the ground very weak. Therefore, the tracking process of the GNSS receiver is confronted with intractable problems in maintaining high sensitivity and alleviating the interference [5], [6].

In general, each satellite signal is processed independently in different tracking channels of the GNSS receiver, where the delay lock loop (DLL), phase lock loop (PLL), and frequency lock loop (FLL) precisely track the code, carrier phase, and the frequency, respectively. However, there is an acute growth in the scope of the GNSS applications, and the receiver gradually becomes more fragile to the unpredictable environments by applying the traditional baseband processing algorithms. Thus, it is very urgent to enhance the tracking ability and design a more rugged GNSS receiver for flexibly adapting to different navigating situations.

The traditional tracking method is based on the scalar tracking loop accounting for the individually independent tracking channels. The tracking and navigating processors are independent. By contrast, a vector tracking technique altogether processes the baseband and navigation data [6]. The navigation solutions contribute to the tracking processing such that a closed-loop architecture is developed in this way. The vector tracking method reduces the tracking noise of the weak channel, so it is easier to constrain the tracking error residuals within a linear region. Besides, this technique is able to process the multi-channel signals even if one or more incoming signals are blocked. In addition, it is superior in weak- and dynamic-signal tracking and has enhanced anti-jamming performance [7], [8].

Vector tracking has attracted interest from many researchers and has been hotly debated since its prototype, i.e., vector delay lock loop (VDLL), was firstly presented by Spilker [6]. For example, it has been proved that vector tracking could lift the navigation performance of the receiver in high-dynamic and weak situations, respectively [9], [10]. It has also demonstrated that the vector tracking algorithms performed high spoofing mitigation and anti-jamming [11], [12]. An open-source vector receiver was introduced to alleviate the multipath interference and the non-light-of-sight (NLOS) reception [13], [14]. An improved vector phase lock loop (VPLL) was presented based on the double-difference carrier positioning algorithm to efficiently reduce the clock drift and atmospheric delay errors [15]. Besides, the vector tracking was verified to perform well in ionospheric scintillation mitigation [16]. Bhattacharyya elaborated on the receiver autonomous integrity monitoring (RAIM) performance in the vector receiver and provided core mathematical and theoretical models corresponding to the vector tracking model [17]. Lashley investigated the various architectures of the vector tracking technique in detail, where a vector delay/frequency lock loop (VDFLL) was deeply

Yiran Luo and Naser El-Sheimy are with the Department of Geomatics Engineering, University of Calgary, Calgary, AB T2N 1N4, Canada (e-mail: yiran.luo@ucalgary.ca; elsheimy@ucalgary.ca).
You Li is with the State Key Laboratory of Information Engineering in Surveying, Mapping and Remote Sensing (LIESMARS), Wuhan University, Wuhan, China, 430079 (e-mail: liyou@whu.edu.cn).
Jin Wang is with the College of Geodesy and Geomatics, ShanDong University of Science and Technology, Qingdao, China, 266590 (e-mail: wangjin@sdust.edu.cn).

researched. It could efficiently process weak signals. [18]. Then, the VDFLL technique was validated to improve the high-precision positioning performance in a harsh environment [1], [19].

Once the number of visible GNSS satellites is sharply degraded, the vector tracking will be less powerful. In this case, the standalone receiver is difficult to provide reliable navigation solutions. The inertial navigation system (INS) is a promising candidate for the GNSS-based navigation system resisting various challenges [20]. The GNSS is sensitive to high dynamics, while the INS inversely adapts very well. Accordingly, the INS is commonly integrated with GNSS for designing the ultra-tight coupling integration navigation system [21]. The loose coupling and tight coupling of GNSS and INS are also frequently used to improve the navigation performance of the integrated system [22]–[24]. Whereas these two do not change the GNSS baseband.

Many recent research works have focused on the ultra-tightly coupled GNSS/INS integration navigation techniques. The researchers from Wuhan University have done many works on a type of cascaded ultra-tight integration algorithm of GNSS and INS and realized it on a hardware prototype [25]–[27]. Their contributions prove that the ultra-tight integration can highly improve the navigation performance of the GNSS receiver in challenging environments. Next, the improved federated ultra-tight coupling GPS/INS and BDS/INS integration architectures were presented. They were verified to have a higher performance in the weak-signal and dynamic situations [28], [29]. An interference detector unit was introduced in an improved ultra-tight coupling receiver to enhance its anti-jamming performance further [7]. In addition, an ultra-tightly coupled GPS/BDS/INS integration algorithm was proposed to improve the navigation robustness in a high-dynamic situation by using both the GPS and BDS information [30], [31]. A sample-wise aiding algorithm was introduced to enhance the dynamic performance of the GPS/INS ultra-tight coupling system as well [32]. This method performs better using the more low-cost inertial measurement unit (IMU). Moreover, the phased array antenna was adopted in an ultra-tight integration system by which the estimation accuracies of positioning and determination were raised, and the anti-jamming ability was enhanced [33].

Previous papers verified that the MEMS/low-cost IMU could improve GNSS receiver performance [34]–[37]. However, the sensors used in these pieces of research still cannot represent the most low-cost ones in the commercial and industrial fields. Recently, an ultra-tightly coupled integration of GNSS and INS based on the consumer-level sensor was verified to help the high-precision positioning for ground vehicle navigation [38]. However, the discrepancies between different types of IMUs and different types of fusing information are not assessed. As the signal reception condition intimately threatens the navigation performance of a GNSS receiver in the challenging ground case, this paper, an upgraded version of [38], focuses on answering whether the smartphone-based consumer-level IMU is sufficient to support the baseband of the GNSS receivers.

Considering the acutely increasing potential of the smartphone-sensor application in navigation and the advantages of vector tracking for the GNSS baseband processor, the main contributions of this work are stated as follows:

1) For the issue of fusing the smartphone IMU data with the deep GNSS baseband information, an extended Kalman filter (EKF) navigator processing the GNSS measurements and their INS predictions is incorporated with the VDFLL such that a hybrid cascaded architecture towards the ultra-tight integration is formed;
2) Then, in order to build an ultra-tight integration platform to realize the proposed fusing algorithm, considering a complete system design from the signal source processing to the positioning, velocity, and timing (PVT) solutions, a GNSS software-defined radio (SDR) receiver is upgraded and integrated with INS mechanization;
3) Finally, confronted with the problem of whether the smartphone sensor can help the GNSS baseband, both tracking and carrier-based positioning performances are assessed based on the presented ultra-tight platform via the real-world kinematic filed test.

This paper is organized as follows: Section II investigates the architecture of the cascaded ultra-tightly coupled integration system; then, the integrated EKF algorithm is introduced in Section III; in Section IV, the field tests are carried out for ground vehicle navigation where a GNSS SDR platform is ultra-tightly integrated with the IMU data, and the results are analyzed and discussed; finally, Section V concludes the paper.

## II. System design

This part will focus on the cascaded ultra-tight integration of GNSS and INS, an upgraded version of the VDFLL. The VDFLL architecture can refer to the authors' previous works [19], [39]. The architecture of a hybrid cascaded ultra-tight integration of the GNSS and INS will be investigated at first; then, the timelines of the working process for the loop filter algorithms will be discussed.

### A. *The architecture of the hybrid cascaded ultra-tightly coupled integration*

In contrast to a centralized vector tracking loop where tracking signals of all channels are processed in a single navigation filter, the cascaded counterpart includes a single pre-filter with each of the channels to estimate tracking errors, respectively [8], [40]. This work will use the cascaded architecture to form an ultra-tight system. Next, in an ultra-tight structure, the integrated EKF solutions, the INS dead reckoning (DR) results, and the outputs from the GNSS baseband processing need to be integrated. Thus, this research adopts a hybrid integration for the ultra-tight system design. The architectures are shown in Fig. 1, where the IMU data aid the GNSS tracking loops, and three tracking approaches (i.e., the scalar tracking with carrier aiding, VDFLL aided by INS DR, and VDFLL aided by both integrated EKF and INS DR) are incorporated. Then, the updating rates for the three tracking approaches are 200 Hz, 5 Hz, and 1 Hz, respectively. It is worth mentioning that clock bias and drift estimations are updated every second in Fig. 1c;

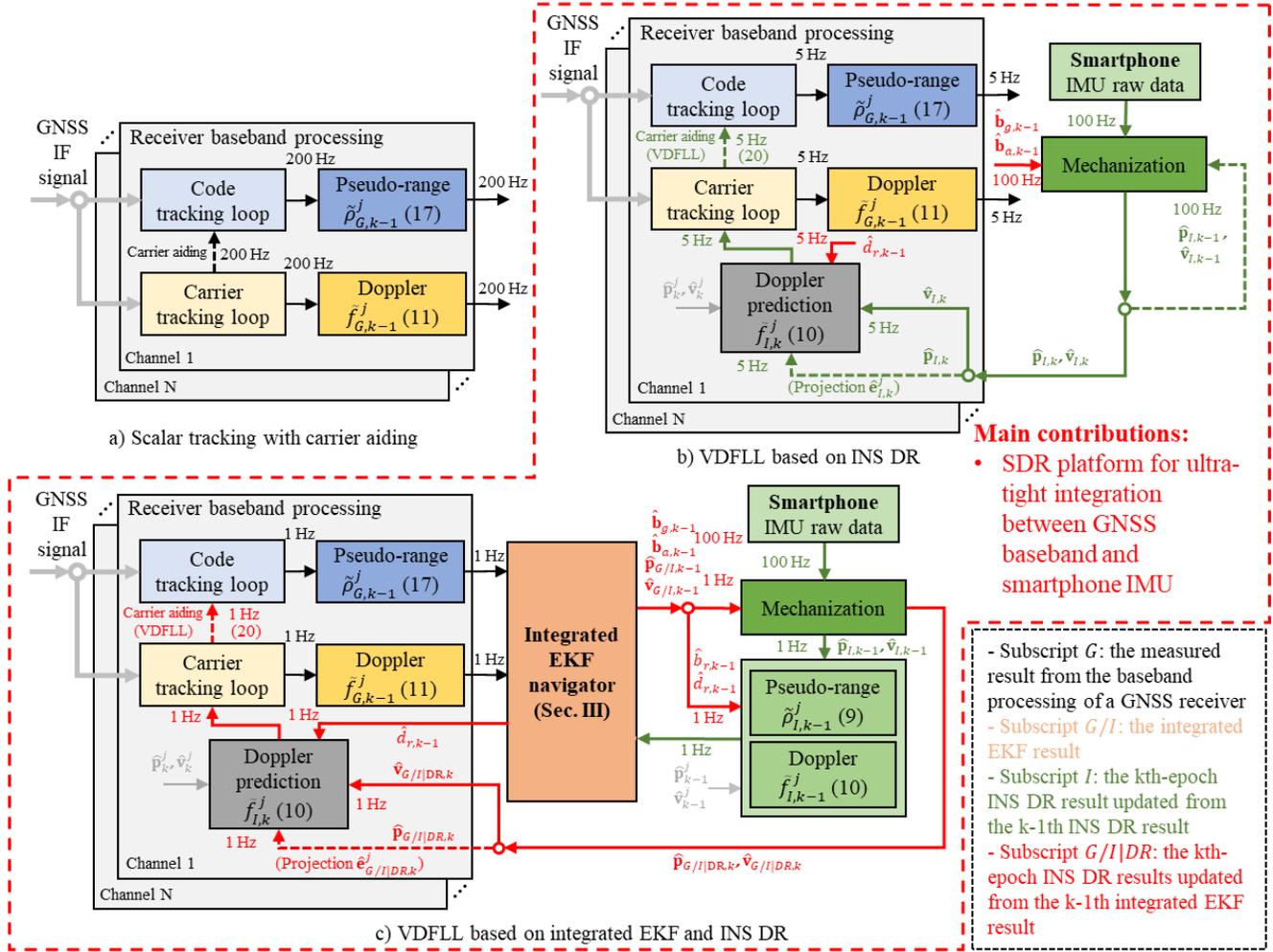

Fig. 1 Hybrid cascaded architectures of the ultra-tightly integrated GNSS receiver aided by the IMU ($k$ is the updating epoch index; $\hat{\mathbf{p}}_k$, $\hat{\mathbf{v}}_k$, $\hat{\mathbf{p}}_k^j$, and $\hat{\mathbf{v}}_k^j$ are the estimated user's position and velocity, the predicted position and velocity of satellite $j$, respectively; $\hat{\mathbf{b}}_{g,k-1}$, $\hat{\mathbf{b}}_{a,k-1}$, $\hat{b}_{r,k-1}$ and $\hat{d}_{r,k-1}$ are the respective estimated gyro, accelerometer bias vector along three axes, estimated clock bias and clock drift; $\hat{\mathbf{e}}_k^j$ is the cosine direction unit vector; red and greens lines correspond to the integrated EKF results and the standalone INS mechanization results, respectively. ).

each group of the updates will be used for one Doppler prediction in Fig. 1c and four such predictions in Fig. 1b.

### B. Baseband tracking algorithm

Detailed architectures of the tracking loops are illustrated in Fig. 2. The traditional DLL and PLL are used to fuse the IMU data. At the same time, the predicted Doppler assists the numerically controlled oscillator (NCO) carrier frequency from the INS DR and the integrated EKF. Finally, the ephemeris from the offline RINEX file is used to compute satellite positions $\mathbf{p}_k^j$ and velocities $\mathbf{v}_k^j$ where the superscript $j$ is the pseudo-random noise (PRN) number of the satellite and the subscript $k$ is the epoch index.

The timeline of how the GNSS tracking, GNSS/INS integrated EKF, and INS DR results are incorporated is shown in Fig. 3. In the system, the updating rate of the integrated EKF is 1 Hz. The INS DR updating rate is 5 Hz for the ultra-tight integration tracking loop. In other words, the code and carrier tracking loops will be aided by the PVT estimations from the integrated EKF once and from the single INS DR four times every second.

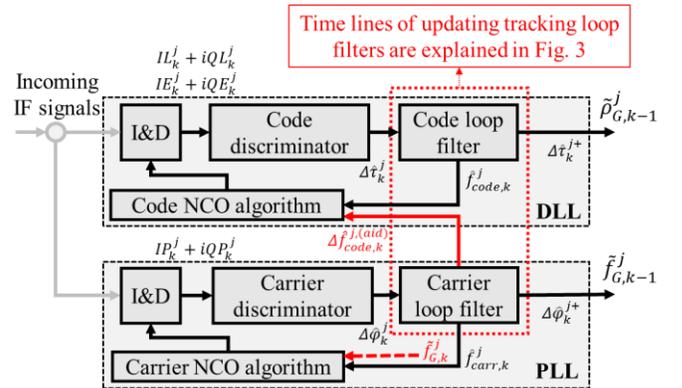

Fig. 2 Details of the tracking loop architectures used for the ultra-tight integration where the denotations can be found in Methodology.

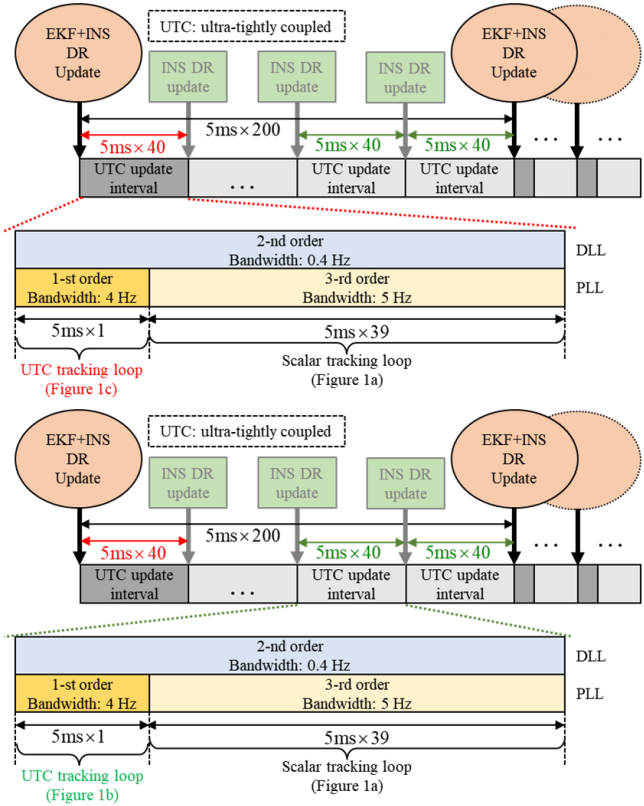

Fig. 3 Timeline of updating tracking loop filters.

Again, as the coherent integration interval is 5 ms and non-coherent integration is not used, the updating rate of the tracking loop is 200 Hz. Therefore, the vector tracking approach is adopted for the first 5 ms of each 200-ms time spanning while the carrier NCO is assisted by the predicted Doppler from the users' PVT solutions. Next, the tracking loop will be modified to the traditional scalar tracking mode for the remaining 39 5-ms intervals over 200 ms. This modification aims to reduce the time-correlated error induced by the cascaded estimators in the tracking loop.

It is worth mentioning that once positioning outliers are detected over the INS DR updating interval, the navigation system will switch to the standalone VDFLL algorithm.

### III. INTEGRATED EKF ALGORITHM

The integrated EKF related to the architecture, as shown in Fig. 1c, is constructed in the Earth-centered Earth-fixed (ECEF) frame to fuse the GNSS measurements and the INS predictions. The state transition equation, observation equation, and noise covariance matrices related to the EKF algorithm will be investigated.

#### A. State transition equation

This section investigates the state transition equation. Compared with the previous work [38], the clock bias and drift errors can be estimated here as the raw GNSS measurements are processed.

The state transition equation is given by

$$\delta \mathbf{x}_k^e = \mathbf{\Phi}_{k,k-1}^e \delta \mathbf{x}_{k-1}^e + \mathbf{w}_{k-1}^e \quad (1)$$

where $\mathbf{w}_{k-1}^e$ is the process noise vector. Then, the state vector is given by

$$\delta \mathbf{x}_k^e = \left[ (\delta \mathbf{\psi}^e)^T, (\delta \mathbf{v}^e)^T, (\delta \mathbf{r}^e)^T, (\delta \mathbf{b}_g)^T, (\delta \mathbf{b}_a)^T, \delta b_r, \delta d_r \right]_k^T \quad (2)$$

where $\delta \mathbf{\psi}^e$ denotes the attitude error vector; $\delta \mathbf{v}^e$ is the velocity error vector; $\delta \mathbf{r}^e$ is the position error vector; $\delta \mathbf{b}_g$ and $\delta \mathbf{b}_a$ stand for the state vectors of gyro and accelerometer bias errors, respectively; $\delta b_r$ is the clock bias error state, while $\delta d_r$ denotes the clock drift error state. Next, the dynamic matrix is given by

$$\mathbf{F}^e = \begin{bmatrix} \mathbf{F}_I & \mathbf{0} \\ \mathbf{0} & \mathbf{F}_G \end{bmatrix}, \mathbf{F}_G = \begin{bmatrix} 0 & 1 \\ 0 & 0 \end{bmatrix},$$

$$\mathbf{F}_I = \begin{bmatrix} -\mathbf{\Omega}_{ie}^e & \mathbf{0}_3 & \mathbf{0}_3 & \hat{\mathbf{C}}_b^e & \mathbf{0}_3 \\ \mathbf{F}_{21}^e & -2\mathbf{\Omega}_{ie}^e & \mathbf{F}_{23}^e & \mathbf{0}_3 & \hat{\mathbf{C}}_b^e \\ \mathbf{0}_3 & \mathbf{I}_3 & \mathbf{0}_3 & \mathbf{0}_3 & \mathbf{0}_3 \\ \mathbf{0}_3 & \mathbf{0}_3 & \mathbf{0}_3 & -\mathbf{\beta}_g & \mathbf{0}_3 \\ \mathbf{0}_3 & \mathbf{0}_3 & \mathbf{0}_3 & \mathbf{0}_3 & -\mathbf{\beta}_a \end{bmatrix} \quad (3)$$

with

$$\mathbf{F}_{21}^e = \left[ -\left( \hat{\mathbf{C}}_b^e \hat{\mathbf{f}}_{ib}^b \right) \times \right], \quad \mathbf{F}_{23}^e = -\frac{2\hat{\mathbf{\gamma}}^e}{r_s(\hat{\phi})} \frac{(\hat{\mathbf{r}}^e)^T}{|\hat{\mathbf{r}}^e|} \quad (4)$$

where $(\cdot) \times$ denotes the operator of the skew-symmetric matrix; $\mathbf{\Omega}_{ie}^e$ is the skew-symmetric matrix of the Earth-rotation vector; $\hat{\mathbf{C}}_b^e$ represents the body-to-Earth-frame coordinate transformation matrix in current epoch; $\hat{\mathbf{f}}_{ib}^b$ is the measurement vector of the accelerometer; $\hat{\mathbf{\gamma}}^e$ represents the gravitational acceleration vector in the ECEF-frame axes; $r_s(\hat{\phi})$ stands for the geocentric radius at the surface which varies with the latitude $\hat{\phi}$ [20]. The gyro bias error and accelerometer bias error are modeled with first-order Gauss-Markov (GM) processes where $\mathbf{\beta}_g$ and $\mathbf{\beta}_a$ are given by

$$\mathbf{\beta}_g = \beta_g \mathbf{I}_3, \quad \mathbf{\beta}_a = \beta_a \mathbf{I}_3 \quad (5)$$

where $\beta_g$ and $\beta_a$ denote the reciprocal of the time constant in the gyro and accelerometer bias error models, respectively; $\mathbf{I}_m$ is the identity matrix with the dimension of $m$. Then, ignoring the superscript of $\mathbf{F}^e$, the transition matrix $\mathbf{\Phi}_{k+1,k}^e$ can be approximated as [41]

$$\mathbf{\Phi}_{k+1,k}^e = e^{\mathbf{F} \cdot \Delta t} \approx \mathbf{I}_{17} + \mathbf{F}\tau + \frac{1}{2}\mathbf{F}^2 \tau^2 + \cdots \quad (6)$$

#### B. Observation equation

The observation equation is provided as

$$\delta \mathbf{z}_k = \mathbf{H}_k^e \delta \mathbf{x}_k^e + \mathbf{v}_k \quad (7)$$

where $\mathbf{v}_k$ is the observation noise vector. Next, the observation vector consists of the pseudorange and Doppler as

$$\delta \mathbf{z}_k = \left[ \tilde{\rho}_I^{(1)} - \tilde{\rho}_G^{(1)}, \cdots, \tilde{\rho}_I^{(n_1)} - \tilde{\rho}_G^{(n_1)}, \tilde{f}_I^{(1)} - \tilde{f}_G^{(1)}, \cdots, \tilde{f}_I^{(n_2)} - \tilde{f}_G^{(n_2)} \right]_k^T \quad (8)$$

where the superscript $n_1$ and $n_2$ correspond to the indexes of the satellite channels related to the measured pseudorange and the measured Doppler, respectively; $\tilde{\rho}_{I,k}$ and $\tilde{f}_{I,k}$ represent the equivalent pseudorange and Doppler predicted from the INS, and they are computed as

$$\tilde{\rho}_{I,k}^j = \left\| \hat{\mathbf{p}}_k^j - \hat{\mathbf{p}}_{G/I|DR,k} \right\| + \left( \hat{b}_{r,k-1} + \hat{d}_{r,k-1} T_{coh} - \hat{b}_k^j \right) + \hat{I}_k^j + \hat{T}_k^j \quad (9)$$

$$\tilde{f}_{I,k}^j = \frac{f_r}{c} \left[ \hat{\mathbf{e}}_{G/I|DR,k}^j \left( \hat{\mathbf{v}}_{G/I|DR,k} - \hat{\mathbf{v}}_k^j \right) + \left( \hat{d}_{r,k-1} - \hat{d}_k^j \right) \right] \quad (10)$$

where $\|\cdot\|$ denotes the $l_2$ norm operator; the subscript $G/I|DR$ represents that the argument is estimated from the previous-epoch integrated EKF and current-epoch INS DR; $\hat{\mathbf{p}}_{G/I|DR,k}$ and $\hat{\mathbf{v}}_{G/I|DR,k}$ denote the estimated user's 3D position and velocity column vector, respectively; $\hat{b}_{r,k-1}$ and $\hat{d}_{r,k-1}$ are the estimated local clock bias and drift from the integrated EKF; $\hat{\mathbf{p}}_k^j$, $\mathbf{v}_k^j$, $\hat{b}_k^j$ and $\hat{d}_k^j$ are the predicted satellite position, velocity, and clock bias and drift from the ephemeris, respectively; $\hat{I}_k^j$ and $\hat{T}_k^j$ are the predicted ionospheric and tropospheric delay errors from Klobuchar and Saastamoninen models, respectively; $\hat{\mathbf{e}}_{G/I|DR,k}^j$ is the cosine direction unit and $\left[ \hat{\mathbf{e}}_{G/I|DR,k}^j, 1 \right]$ is the projection; $T_{coh}$, $f_r$ and $c$ are the coherent integration interval, radio frequency, and the speed of light, respectively

Then, the measured pseudorange $\tilde{\rho}_{G,k}$ and Doppler $\tilde{f}_{G,k}$ from the ultra-tight GNSS receiver will be described.

At first, the Doppler frequency is estimated based on the VDFLL algorithms as

$$\tilde{f}_{G,k}^j = \tilde{f}_{I,k}^j + \frac{1}{T_{coh}} \Delta \hat{\varphi}_{carr,k}^{j+} \quad (11)$$

$$\Delta \hat{\varphi}_{carr,k}^j = \arctan\left( \frac{QP_k^j}{IP_k^j} \right) \quad (12)$$

with

$$QP_k^j \triangleq AR\left( \Delta \tau_k^j \right) \text{sinc}\left( \Delta f_k T_{coh} \right) \sin\left( \pi \Delta f_k T_{coh} + \Delta \varphi_{k,0} \right) \quad (13)$$

$$IP_k^j \triangleq AR\left( \Delta \tau_k^j \right) \text{sinc}\left( \Delta f_k T_{coh} \right) \cos\left( \pi \Delta f_k T_{coh} + \Delta \varphi_{k,0} \right) \quad (14)$$

$$\Delta f_k \triangleq f_{carr,k}^j - \hat{f}_{carr,k}^j \quad (15)$$

$$\hat{f}_{carr,k}^j = f_i + \hat{f}_{G,k-1}^j \quad (16)$$

where $\Delta \hat{\varphi}_{carr,k}^{j+}$ is the estimated carrier phase error passing through the 1-st order loop filter of which the input is $\Delta \hat{\varphi}_{carr,k}^j$ and $\arctan(\cdot)$ is a two-quadrant arctangent discriminator; $QP_k^j$ and $IP_k^j$ are the prompt quadrature and in-phase baseband signal; $A$ is the signal amplitude; $R(\cdot)$ is the auto-correlation function; $\text{sinc}(\cdot)$ is the normalized sinc function and it satisfies $\text{sinc}(x) \triangleq (\pi x)^{-1} \sin(\pi x)$; $\Delta \tau_k^j$ is the code phase error between the replicated and the incoming code signals; $\Delta f_k$ is the frequency error between the replicated frequency $\hat{f}_{carr,k}^j$ and the incoming one; $f_i$ is the intermediate frequency; $\Delta \varphi_0$ is the initial carrier phase error of each updating interval.

Secondly, how to extract the measured pseudorange will be discussed. And it can be computed by

$$\tilde{\rho}_{G,k}^j = c \left( \hat{t}_{r,k} - \frac{1}{f_c} \left( \hat{\tau}_{code,k}^j + \hat{f}_{code,k}^j T_{coh} \right) \right) \quad (17)$$

where $\hat{\tau}_{code,k}^j$ is the remaining code phase terms in chips which can be obtained from the measured week second, the frame, the bit, and the integral code chip information at the local time $\hat{t}_{r,k}$ in seconds [42]; $\hat{f}_{code,k}^j$ is the code NCO frequency estimated from

$$\hat{f}_{code,k}^j = f_c + \hat{f}_{code,dop,k}^j \quad (18)$$

with

$$\hat{f}_{code,dop,k}^j = \Delta \hat{f}_{code,k}^{j,(aid)} + \frac{1}{T_{coh}} \Delta \hat{\tau}_k^{j+} \quad (19)$$

$$\Delta \hat{f}_{code,k}^{j,(aid)} = -\frac{f_c}{f_r} \tilde{f}_{G,k}^j \quad (20)$$

where $f_c$ is the code frequency; $\Delta \hat{\tau}_k^{i+}$ is the output of the 2-nd order loop filter, and its input $\Delta \hat{\tau}_k^i$ is estimated from the noncoherent-early-minus-late-power discriminator, that is,

$$\Delta \hat{\tau}_k^i = \frac{1}{2} \frac{\left( IE_k^j \right)^2 + \left( QE_k^j \right)^2 - \left( IL_k^j \right)^2 - \left( QL_k^j \right)^2}{\left( IE_k^j \right)^2 + \left( QE_k^j \right)^2 + \left( IL_k^j \right)^2 + \left( QL_k^j \right)^2} \quad (21)$$

where $IE_k^j$, $QE_k^j$, $IL_k^j$, and $QL_k^j$ are the early in-phase, quadrature and late in-phase, quadrature baseband signals, respectively.

At last, the observation matrix which combines the state vector with the observation vector is given by

$$\mathbf{H}_k^e = \begin{bmatrix} \mathbf{0}_{n_1 \times 3} & \mathbf{0}_{n_1 \times 3} & -\mathbf{H}_1 & \mathbf{0}_{n_1 \times 6} & \mathbf{I}_{n_1 \times 1} & \mathbf{0}_{n_1 \times 1} \\ \mathbf{0}_{n_2 \times 3} & -\mathbf{H}_2 & \mathbf{0}_{n_2 \times 3} & \mathbf{0}_{n_2 \times 6} & \mathbf{0}_{n_2 \times 1} & \mathbf{I}_{n_2 \times 1} \end{bmatrix}_k \quad (22)$$

where $\mathbf{H}_1$ and $\mathbf{H}_2$ denote the direction cosine matrices of pseudorange and Doppler measurements, respectively.

The Doppler measurement $\tilde{f}_{G,k}^j$ estimated from the prompt carrier NCO is in an instantaneous state. It differs from the standard tightly coupled algorithm, where the Doppler measurement is an averaging and smoothing value over the interval between two epochs. As the updating rate of the tracking loop is high, the loop requires a prompt response to the signal dynamics. Using the instantaneous Doppler value can enhance the integrated EKF to respond to the dynamics of the user and the clock.

In summary, the mathematical models about how the GNSS baseband tracking data are integrated with the INS DR solutions and how the integrated solutions assist the GNSS baseband were introduced.

### C. Noise covariance matrices

The developments of the process and measurement

6covariances matrices are introduced in this part.

The standard deviations related to the gyro and accelerometer bias errors are built following the first-order GM process as mentioned earlier. In this case, the process noise matrix can be approximated as [41]

$$\mathbf{Q} \approx \begin{bmatrix} \mathbf{Q}_1 & \mathbf{0}_{15 \times 2} \\ \mathbf{0}_{2 \times 15} & \mathbf{Q}_2 \end{bmatrix} \Delta t \quad (23)$$

with

$$\mathbf{Q}_1 = \mathrm{diag}\left(\begin{bmatrix} \sigma_{gwn}^2 & \sigma_{awn}^2 & 0 & 2\beta_g \sigma_{gb}^2 & 2\beta_a \sigma_{ab}^2 \end{bmatrix}\right) \otimes \mathbf{I}_3$$

$$\mathbf{Q}_2 = \mathrm{diag}\left(\begin{bmatrix} \sigma_{tb}^2 & \sigma_{td}^2 \end{bmatrix}\right)$$

where $\sigma_{gwn}$, $\sigma_{gb}$, $\sigma_{awn}$ and $\sigma_{ab}$ denote the power amplitude of the gyro bias, gyro drift, accelerometer bias, and accelerometer drift, respectively; $\sigma_{tb}$ and $\sigma_{td}$ stand for the respective standard deviation of the clock bias and drift error; $\Delta t$ is the updating interval of the EKF, which is one second in this work.

As a two-quadrant arctangent and a noncoherent-early-minus-late-power discriminator are used for discriminating the carrier phase error and code error, respectively, in the loop filter, the measurement noise matrix is given by

$$\mathbf{R} = \begin{bmatrix} \left(\kappa_{\Delta\rho}\right)^2 \mathbf{R}_r & \mathbf{0}_{n_1 \times n_2} \\ \mathbf{0}_{n_2 \times n_1} & \left(\kappa_{\Delta\dot{\rho}}\right)^2 \mathbf{R}_f \end{bmatrix}_{(n_1+n_2) \times (n_1+n_2)} \quad (24)$$

with

$$\mathbf{R}_r = \mathrm{diag}\left(\begin{bmatrix} \left(\sigma_{dll}^{(1)}\right)^2 & \cdots & \left(\sigma_{dll}^{(n_1)}\right)^2 \end{bmatrix}\right)$$

$$\mathbf{R}_f = \mathrm{diag}\left(\begin{bmatrix} \left(\sigma_{pll}^{(1)}\right)^2 & \cdots & \left(\sigma_{pll}^{(n_2)}\right)^2 \end{bmatrix}\right) \quad (25)$$

where the carrier and code thermal noise jitters, i.e., $\sigma_{pll}$ and $\sigma_{dll}$, are reckoned as the primary error sources for the loop filter outputs, and the equations can be found in the references [5], [43]. The compensated factors, i.e., $\kappa_{\Delta\rho}$ and $\kappa_{\Delta\dot{\rho}}$, are selected from empirical experiences to make a trade-off between the pseudorange or Doppler measurement performance and the unexpected error such as the one caused by multipath interferences [44].

Finally, the prediction and update for the state vector are achieved based on the EKF recursive formula [45]. The loop filter algorithms can refer to the previous reference [5] such that they will not be mentioned in this work.

## IV. FIELD TESTS AND RESULTS

This section assesses the navigation and positioning performances of the ultra-tight coupled integration systems when a consumer-level IMU is used. The raw data was collected from the devices installed on a ground vehicle moving around the University of Calgary campus. In this situation, the received GNSS signals were frequently degraded by the surrounding blockages, e.g., tall buildings and foliage. The phenomenon acutely jeopardized the capability of providing high-quality measurements of the standalone GNSS receiver. Therefore, the proposed GNSS SDR platform is used to process the intermediate frequency (IF) GNSS data. An open-source package program, RTKLIB, is used to compute the high-precision RTK positioning solutions [46].

### A. Experimental setup

The experimental equipment is set up on the roof of the ground vehicle, as shown in Fig. 4. A NovAtel antenna is used to receive the GNSS signals through a GNSS receiver frond-end and a low-end U-Blox NEO-M8T receiver. They share the same incoming RF signals through a splitter. In addition, the high-performance Trimble R10 receiver with an independent antenna provides the reference positioning solution. Two types of IMU data, which are, respectively, collected from a low-cost MEMS GPS/INS integration system, Moog Crossbow Nav 440, and LSM6DSM, are contained in the experiment as the source of the mechanization of the INS, where a consumer-level LSM6DSM chip from STMicroelectronics is integrated into the Android-based HUAWEI Mate 9 smartphone. The gyro bias stability values for these two IMUs are 10 deg/h and 7200 deg/h, respectively, while the accelerometer bias stabilities are 1 mg and 40 mg, respectively. Finally, the raw data rates of Nav 440 and LSM6DSM IMUs are respective 100 Hz and 200 Hz.

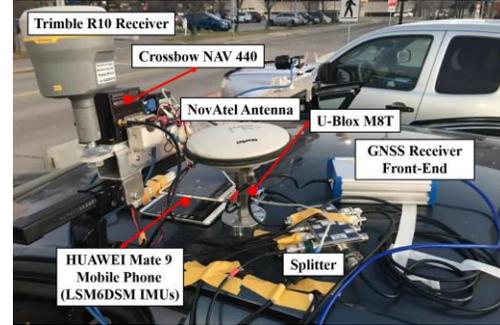

Fig. 4 Setup for the field test towards the ground vehicle navigation based on the ultra-tightly coupled GNSS/INS integration technique.

It is worth emphasizing that the cascaded ultra-tight integration systems are classified into pseudorange-state-based and position-state-based architectures. Meanwhile, the former is based on the EKF model proposed in this work, and the latter can refer to the authors' previous work [38].

Four ultra-tight integration methods are tested as the different comparisons based on combining two IMUs and two architectures. Some core information of the four ultra-tight integration algorithms is summarized in Table I.

TABLE I
LIST OF THE EXPERIMENTAL ULTRA-TIGHT INTEGRATION ALGORITHMS.

| Algorithm | IMU source | Measurement vector for the integrated EKF | Model of the measurement vector |
|---|---|---|---|
| LSM6DSM-LC | LSM6DSM | 3D position and velocity errors | (5) in [38] |
| LSM6DSM-TC | LSM6DSM | Pseudorange and Doppler errors | (8) |
| Nav440-LC | Crossbow Nav 440 | 3D position and velocity errors | (5) in [38] |
| Nav440-TC | Crossbow Nav 440 | Pseudorange and Doppler errors | (8) |

Note: In this work, "LC" and "TC" correspond to the fusing methods of the observations (the former is for position and velocity, and the latter is for pseudorange and Doppler) in the ultra-tight coupling algorithm; they do not mean "loose coupling" and "tight coupling" algorithms.

As mentioned earlier, the IMU raw data from the smartphone is super coarse. Thus, it is interesting to know if such consumer-level inertial sensors contribute to the ultra-tight integration system in ground vehicle navigation. Furthermore, as the Nav 440 is a commercial GPS/INS integration system, it is meaningful to compare the smartphone IMU with the commercial device.

On the other hand, in comparison with position-based fusion, pseudorange-based fusion naturally removes the unexpected errors produced by the receiver navigator. So, it is attractive to figure out if the pseudorange-state-based ultra-tight architecture can exceed the position-state-based one, especially in the kinematic situation where signal blockage and attenuation frequently occur. Finally, the standalone VDFLL, which the IMU does not assist, is also tested as a comparison, and the algorithm can refer to the authors' previous work [19].

The GNSS SDR processes the GPS L1 C/A signal in this test, and the experiment lasted for around 820 seconds. The corresponding parameter settings for the experiment are provided in Table II.

TABLE II
PARAMETER CONFIGURATIONS OF THE GNSS SDR AND THE IMUS FOR THE ULTRA-TIGHT INTEGRATION EXPERIMENT.

| Parameter | Value |
| --- | --- |
| Signal Type | GPS L1 C/A |
| Sampling Rate | 10.125 MHz |
| Coherent Integration Time | 5 ms |
| 1st-order PLL Bandwidth | 4 Hz |
| 2nd-order DLL Bandwidth | 0.4 Hz |
| Code Phase Discriminator | Noncoherent-early-minus-late-power discriminator |
| Carrier Phase Discriminator | Costas two-quadrant arctangent discriminator |
| Early-Late Spacing of Code Discriminator | 0.2 chips |

The sky plot for the available satellites during the real-world experiments is illustrated in Fig. 5. Maximum of ten satellites are processed by the GNSS SDR, i.e., SV3, SV6, SV9, SV14, SV16, SV22, SV23, SV25, SV26, and SV31, where SV25 embraces an extremely low elevation angle which is below ten degrees. Then, the carrier-to-noise density ratio ($C/N_0$) estimates are plotted in Fig. 6. It is worth noting that the $C/N_0$ estimated from the four ultra-tight GNSS SDRs are very close to the results from the commercial U-Blox receiver. Thus, the proposed SDR continuously tracked the satellite signals over the field test. It is the prerequisite for the following discussions.

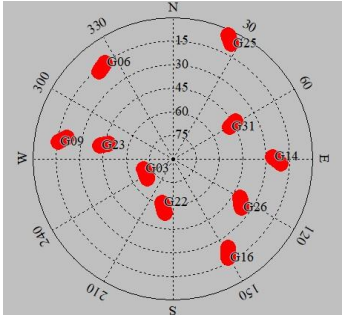

Fig. 5 Sky plot of the used GPS satellites.

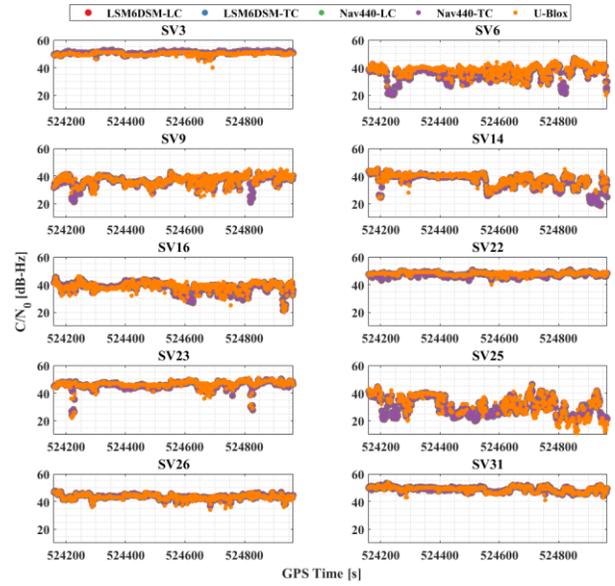

Fig. 6 Curves of $C/N_0$ estimations of the different GNSS SDRs.

### B. Integrated EKF results

The trajectories estimated from the single-point positioning algorithm of the VDFLL-based standalone SDR and the ultra-tight SDRs are illustrated in Fig. 7 (top) and Fig. 8 (top), respectively. It should be noted that all the results associated with the ultra-tight coupling algorithms could always produce the correct trajectory estimations. However, the standalone VDFLL-based SDR has failed to offer reliable positioning results since the middle stage of the testing. The results prove that the ultra-tight integration has more robust stability and continuity than vector tracking.

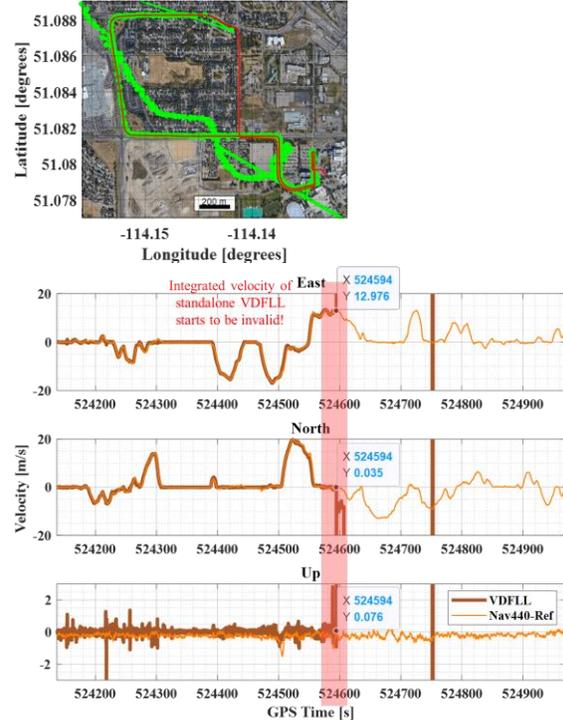

Fig. 7 Integrated trajectory (top) and velocity (bottom) estimation of the standalone VDFLL-based GNSS SDR; in the trajectory picture, green points correspond to estimation results, and the red line corresponds to the reference trajectory.

Next, the velocity curves estimated from the VDFLL and the ultra-tight integration algorithms are also shown in Fig. 7 (bottom) and Fig. 8 (bottom), where "Nav440-Ref" denotes the integrated velocity from the Crossbow Nav 440 navigation system as the reference. Again, it can be found that ultra-tight integration velocities are consistent with the reference values.

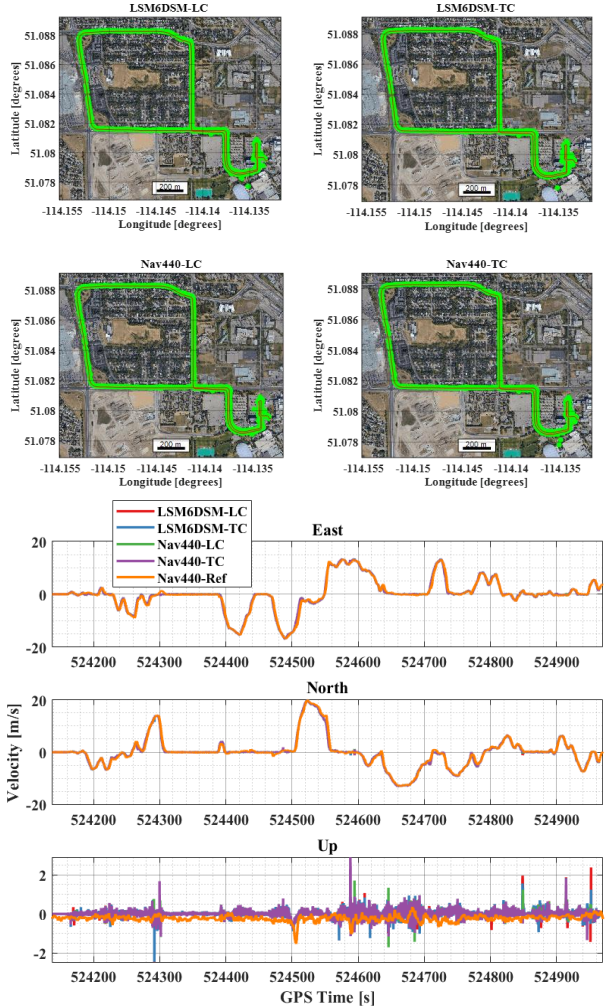

Fig. 8 Integrated trajectory (top) and velocity (bottom) estimations of the four types of ultra-tightly coupled GNSS/INS integration algorithms; in the trajectory picture, green points correspond to estimation results, and the red line corresponds to the reference trajectory.

### C. Tracking results

The mean and standard deviation (STD) values of the carrier phase error over the entire testing time (from GPS time 524160s to 524970s) are computed in Fig. 9. Inferred from the STDs of weak tracking channels (e.g., SV6, SV9, and SV25), the pseudorange-state-based ultra-tight integration with a higher-grade IMU is superior in reducing the bias error in the carrier phase tracking. Besides, for the pseudorange-state-based ultra-tight algorithm, the higher-grade IMU is more efficient in alleviating the tracking error even if the incoming signal power is not weak (seeing the estimating results of SV3, SV22, SV23, and SV31). However, the results show that the position-state-based architecture does not follow this rule. Furthermore, the results also prove that different ultra-tight integration algorithms have little influence on the STDs of carrier phase errors within a short-term GNSS updating interval (1 second).

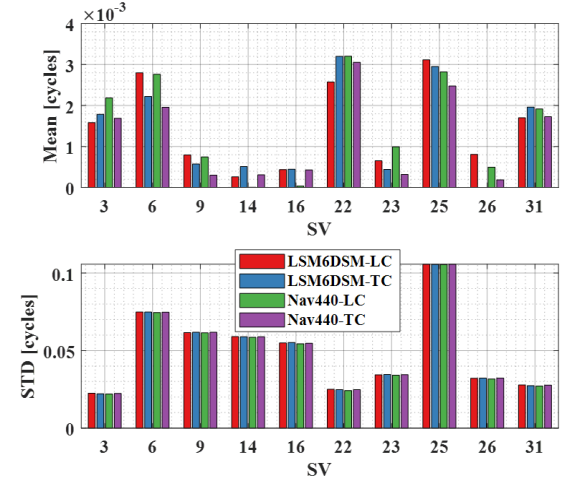

Fig. 9 Means and STDs of the discriminated carrier phase errors.

In the VDFLL, the SDR starts to be invalid at the GPS time of around 524594 s as the PVT solutions have not been reliable since that time, as shown in Fig. 7.

Case studies of the tracking status for SV6 and SV23 are shown in Fig. 10 to Fig. 14 over this period. The prompt in-phase ($I_p$) and quadrature ($Q_p$) samples, carrier, and code discriminating outputs are tested and plotted in the experiment. Both tracking channels with low (SV6) and medium (SV23) elevation angles based on the VDFLL fail to modulate the data code at around 524594.5 s. As shown in Fig. 11, since then, the code and carrier discriminators of SV6 have produced much larger errors than those estimated from the ultra-tight SDRs.

Fig. 12 shows that the signal power of SV23 is scattered to in-phase and quadrature branches, leading to the failure of the data bit extraction from the in-phase channel. At first, it can be inferred from Fig. 13 that the carrier tracking loop of the VDFLL is affected by an incorrect Doppler prediction. Then, it causes biased estimates of the code discriminator. The ultra-tight integration tracking loop is assisted by the IMU such that the Doppler estimation is more robust and the signal maintains locked.

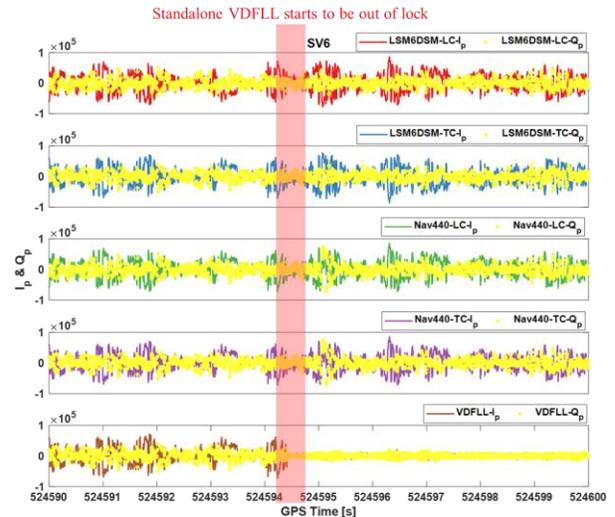

Fig. 10 Case study of prompt in-phase and quadrature values of SV6.

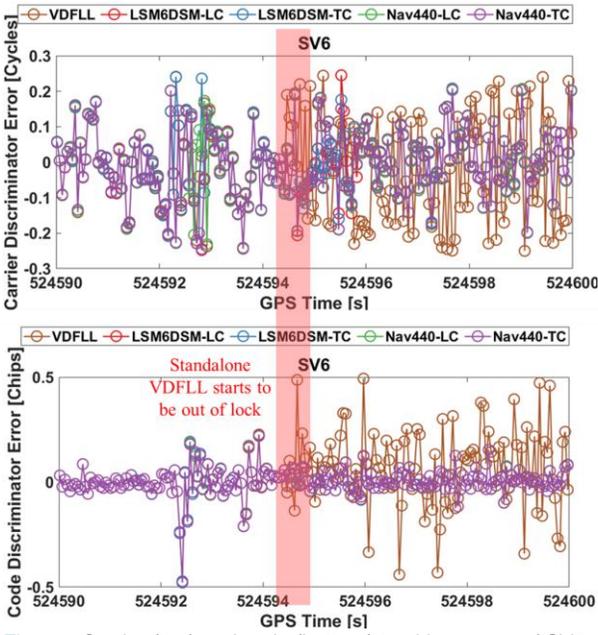

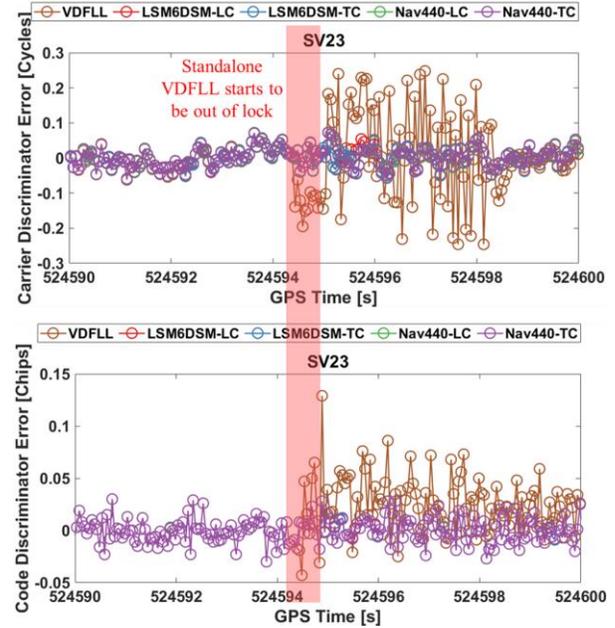

the averaging in-phase value of around $1\times10^5$ is due to the code frequency offset. The results are consistent with the previous discussions in Fig. 12 and Fig. 13.

Fig. 11 Carrier (top) and code (bottom) tracking errors of SV6.

Fig. 13 Carrier (top) and code (bottom) tracking errors of SV23.

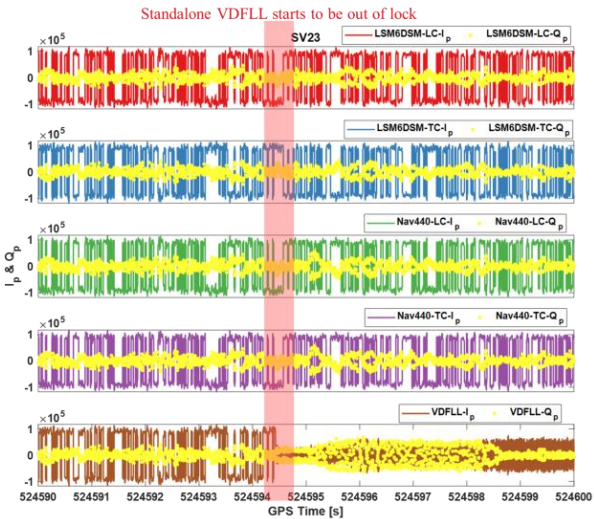

Fig. 12 Case study of prompt in-phase and quadrature values of SV23.

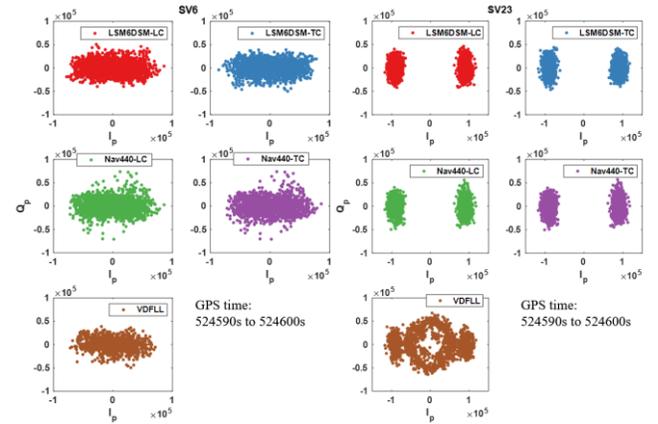

Fig. 14 2D plots of in-phase and quadrature values of SV6 (left) and SV23 (right).

Two-dimensional (2D) plots in terms of the in-phase and quadrature components are shown in Fig. 14. The received signal from SV6 is weak, seeing Fig. 5 and Fig. 6. Thus, it is reasonable that the in-phase and quadrature correlations in this channel are not adequately separated in this 2D plane. Due to the results above, the signal from the VDFLL is lost over this period. So more of the in-phase components are closer to the zero coordinate.

As the signal of SV23 has higher $C/N_0$ performance than the SV6, the signal amplitude is larger, and the groups of in-phase and quadrature components are less close to each other except for the ones from the VDFLL receiver. Then, Fig. 14 (right) also illustrates that a small circle appears in the VDFLL compared to the ultra-tight integration algorithms. Two explanations can be provided: first, the circle is caused by the carrier frequency offset between the received and the local signals. Second, the smaller radius of the circle compared with

As the differences among the four ultra-tight integration algorithms are difficult to quantify in the tracking results, the carrier-based positioning results based on the measurements produced by the four SDRs are subsequently assessed for further comparisons.

### D. RTK positioning results

The RTKLIB is used to implement the RTK algorithm to the carrier measurements in post-processing. The elevation mask is set as ten degrees, and the detailed parameter settings for this software are provided in the authors' previous publication [1].

At first, the RTK error curves are illustrated in Fig. 15, where "U-Blox" corresponds to the RTK results produced by the U-Blox NEO-M8T receiver. At the initial stage of the experiment, the scalar algorithm was applied to enable the receiver to converge and maintain sufficient stability. Then, the ultra-tight coupling algorithm was carried out at the GPS time of 524160 s. Compared with the U-Blox receiver using the scalar algorithm, the EKF in the ultra-tight receiver costs extra time to get the measurements converged. The initial convergence emerged at the epoch around 524350 s.

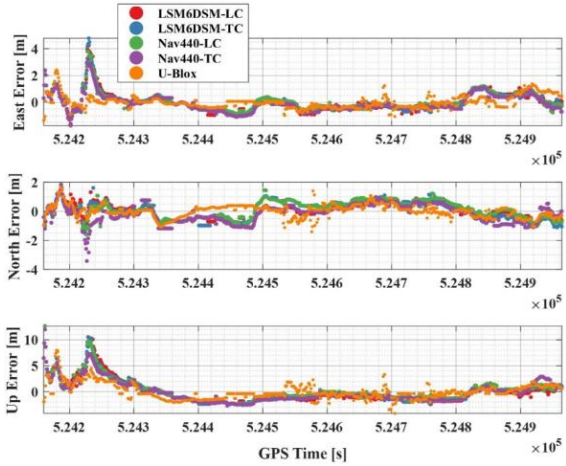

Fig. 15  RTK positioning errors in the field test.

The bars in Fig. 16 manifest the corresponding RTK positioning root-mean-square errors (RMSEs) within the complete and converged data. At first, there are no distinctive differences among the four algorithms after the EKF converges. However, the pseudorange-state-based integrated or higher-grade IMU algorithm can make the ultra-tight integration receiver perform better at the initial stage of the RTK positioning process. In summary, the higher-grade sensor used in the ultra-tight coupling GNSS receiver could slightly strengthen the rapid convergence at the initial time. However, at the same time, it has no distinctive effectiveness in terms of accuracy after the integration algorithm converges.

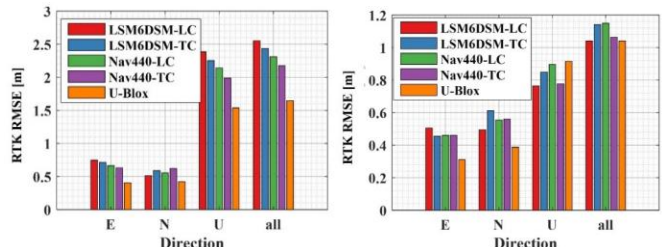

Fig. 16  RMSE of the RTK positioning errors corresponded to the full-datum segment (left) and the converged segment (right).

The summary of the RTK solution type and the visible satellite number is listed in Table III. The higher-grade IMU can enhance the ability of the ultra-tight receiver to fix the ambiguity of the RTK positioning process. For example, the success rates of fixed solutions are computed as 18.6% and 14.3% for Nav440-LC and LSMDSM-LC, respectively, while Nav440-TC and LSM6DSM-TC correspond to 25.4% and 24.2%, respectively. The success rates of the higher-grade IMU are higher than the lower-grade one. One interesting result is that the performance is solely slightly improved when using the pseudorange measurements in the integrated EKF. Besides, the pseudorange-state-based algorithm also outperforms the position-state-based one in fixing the carrier phase positioning. More specifically, the success rates associated with LSM6DSM-LC and Nav440-LC are respectively 14.3% and 18.6%, while LSM6DSM-TC and Nav440-TC are respectively 9.9% and 6.8% higher.

TABLE III
SATELLITE NUMBER AND SOLUTION TYPE OF THE RTK RESULTS.

| Receiver Type | Mean # of Satellite | Total Point | Fixed Point | Float Point | % of fixed Solutions |
|---|---|---|---|---|---|
| LSM6DSM-LC | 8.807 | 818 | 117 | 701 | 14.3% |
| LSM6DSM-TC | 8.797 | 819 | 198 | 621 | 24.2% |
| Nav440-LC | 8.802 | 819 | 152 | 667 | 18.6% |
| Nav440-TC | 8.811 | 819 | 208 | 611 | 25.4% |

It is worthwhile to note that a higher-grade IMU only marginally improves the success rate in our research. At first, we know that an initial error in the integrated EKF represents the statistics of the navigation errors after each GNSS updating timestamp. Again, the navigation error independent of the motion dynamics determines if the low-cost IMU can be available for ultra-tight integration. It proves that the tracking error is primarily from the initial velocity and attitude errors of the IMU within the 1-s short-term GPS updating interval [35].

According to the discussions above, the initial errors are very close to each other no matter whether a higher-grade IMU is adopted in this work, as they are mainly determined by the inherent characteristics of the GNSS receiver baseband processors and the geometry distributions of visible satellites. It can also be explained that the error sources from the IMU raw data (e.g., gyro/accelerometer bias stability and angle/velocity random walk) have little influence on the tracking error of the ultra-tight navigation system when the GNSS receiver provides a relatively high data updating rate (e.g., 1 Hz).

Another minor factor is that the tracking error related to the IMU error sources is more dependent on the accelerometer than the gyro sensors. Once a GNSS receiver deeply integrates with a consumer-level IMU except for a higher-grade one, it is more tolerable for the accelerometer error than the gyro.

Therefore, it is reasonable that the navigation system performance is only slightly improved when the consumer-level IMU is replaced by a higher grade one in the ultra-tightly coupled GNSS/INS integration.

## V. CONCLUSIONS

We proposed a cascaded ultra-tightly integrated GNSS/IMU navigation algorithm for the kinematic ground vehicle navigation to evaluate how the smartphone IMU contributes to the GNSS receiver baseband. Four ultra-tight integration algorithms, i.e., LSM6DSM-LC, LSM6DSM-TC, Nav440-LC, and Nav440-TC, were tested based on the GNSS SDR in the real-world experiments where the standalone VDFLL-based algorithm was also tested as a comparison. Conclusions have been drawn from the results as follows:

1) For the tracking performance: a) all the cascaded ultra-tight integration algorithms outperformed the VDFLL algorithm in tracking, especially when the incoming signal power was relatively weak; b) for the pseudorange-state-based cascaded ultra-tight algorithm, the carrier tracking errors aided with the higher-grade IMU were smaller than the results of the smartphone IMU, and it also performed better than the position-state-based one in weak situations;

2) For the measurement quality: a) the higher-grade IMU could marginally accelerate the convergence at the initial stage of RTK positioning; b) the success rate of fixed solutions from the pseudorange-state-based cascaded ultra-tight receivers performed much better than the position-state-based ones. However, it was only slightly improved by the higher-grade IMU algorithm compared with the smartphone IMU.

Overall, the GNSS baseband performance aided with the smartphone IMU can be close to that of a higher-grade IMU by taking advantage of the ultra-tight integration algorithms. Therefore, it shows the potential of enhancing the GNSS-based navigation systems with only a few dollars of extra hardware in the future commercial market.

12[36] B. Liu, X. Zhan, and M. Liu, "GNSS/MEMS IMU ultra-tightly integrated navigation system based on dual-loop NCO control method and cascaded channel filters," *IET Radar, Sonar Navig.*, vol. 12, no. 11, pp. 1241–1250, 2018.

[37] B. Liu, Y. Gao, Y. Gao, K. Tong, and M. Wu, "Non-consecutive GNSS signal tracking-based ultra-tight integration system of GNSS/INS for smart devices," *GPS Solut.*, vol. 26, no. 3, Jul. 2022.

[38] Y. Luo *et al.*, "Assessment of Ultra-Tightly Coupled GNSS/INS Integration System towards Autonomous Ground Vehicle Navigation Using Smartphone IMU," in *2019 IEEE International Conference on Signal, Information and Data Processing (ICSIDP)*, 2019, pp. 1–6.

[39] Y. Luo, L.-T. Hsu, Y. Xiang, B. Xu, and C. Yu, "An Absolute-Position-Aided Code Discriminator Towards GNSS Receivers for Multipath Mitigation," in *Proc. ION GNSS 2021, Institute of Navigation, St. Louis, Missouri, USA, Sep 20-24*, 2021, pp. 3772–3782.

[40] M. G. Petovello, C. O'Driscoll, and G. Lachapelle, "Carrier Phase Tracking of Weak Signals Using Different Receiver Architectures," in *Proceedings of ION NTM 2008, Institute of Navigation*, 2008, pp. 781–791.

[41] R. G. Brown and P. Y. C. Hwang, *Introduction to random signals and applied Kalman filter with Matlab exercises*, 4th ed. Hoboken: Wiley, 2012.

[42] A. J. Van Dierendonck, "GPS Receivers," in *Global Positioning System: Theory and Applications, Volume 1*, B. W. Parkinson, J. J. Spilker Jr, P. Axelrad, and P. Enge, Eds. Washington D.C.: American Institute of Aeronautics and Astronautics, Inc., 1996.

[43] Y. Luo, J. Li, C. Yu, J. Wang, and N. El-Sheimy, "Performance of Carrier-Based High-Precision Positioning Aided by a Vector Tracking Technique for Land Vehicle Navigation," in *The 11th International Conference on Mobile Mapping Technology (MMT 2019), Shenzhen, China*, 2019, pp. 283–290.

[44] L.-T. Hsu, P. D. Groves, and S.-S. Jan, "Assessment of the Multipath Mitigation Effect of Vector Tracking in an Urban Environment," in *Proceedings of ION Pacific PNT 2013, Institute of Navigation, Honolulu, Hawaii, USA*, 2013, pp. 498–509.

[45] R. Faragher, "Understanding the Basis of the Kalman Filter Via a Simple and Intuitive Derivation [Lecture Notes]," *IEEE Signal Process. Mag.*, vol. 29, no. 5, pp. 128–132, Sep. 2012.

[46] T. Takasu and A. Yasuda, "Development of the low-cost RTK-GPS receiver with an open source program package RTKLIB," in *Proceedings of the International symposium on GPS/GNSS, Jeju, Korea*, 2009, pp. 4–6.